\documentclass[aps,twocolumn,floatfix,showpacs]{revtex4}
\usepackage{graphicx}
\def \ds {\displaystyle}

\def \vS {{\bf S}}

\def \vQ {{\bf Q}}

\def \vR {{\bf R}} 
 
\def \vk {{\bf k}}
\def \vv {{\bf v}}
\def \mB {\mu_{\rm B}}

\def \Det {{\rm Det}}
\def \uM {\underline{M}}
\def \uudd {\uparrow \uparrow \downarrow \downarrow }
\def \uuudd {\uparrow \uparrow \uparrow \downarrow \downarrow}
\def \uudud {\uparrow \uparrow \downarrow \uparrow \downarrow}

\begin{document}

\title{The Collinear Magnetic Phases of the Geometrically-Frustrated
Antiferromagnet CuFeO$_2$:  The Importance of Stacking}
\author{R.S. Fishman,$^1$ F. Ye,$^2$  J.A. Fernandez-Baca,$^{2,3}$
J.T. Haraldsen,$^1$ and T. Kimura$^4$}
\affiliation{$^1$Materials Science and Technology Division, Oak
Ridge National Laboratory, Oak Ridge, TN 37831, USA}
\affiliation{$^2$Neutron Scattering Science Division, Oak Ridge
National Laboratory, Oak Ridge, TN 37831, USA}
\affiliation{$^3$Department of Physics and Astronomy, The University
of Tennessee, Knoxville, Tennessee 37831, USA}
\affiliation{$^4$Division of Materials Physics, Graduate School of
Engineering Science, Osaka University, Toyanaka, Osaka, 560-8531,
Japan}

\begin{abstract}
The correct stacking of hexagonal layers is used to obtain accurate
estimates for the exchange and anisotropy parameters of the
geometrically-frustrated antiferromagnet CuFeO$_2$.  Those
parameters are highly constrained by the stability of a collinear
metamagnetic phase between fields of 13.5 and 20 T.  Constrained
fits of the spin-wave frequencies of the collinear $\uudd $ phase
below 7 T are used to identify the magnetic unit cell of the
metamagnetic $\uuudd $ phase, which contains two hexagonal layers
and 10 Fe$^{3+}$ spins.
\end{abstract}


\maketitle

Because of their rich magnetic phase diagrams,
geometrically-frustrated antiferromagnets have long occupied an
important place in condensed-matter physics \cite{Diep04}.  The
antiferromagnetic interactions between the Fe$^{3+}$ spins of
CuFeO$_2$ are geometrically frustrated within each hexagonal plane
since no spin configuration can simultaneously minimize the coupling
energies of all three neighbors around an equilateral triangle.
Unlike for many geometrically-frustrated antiferromagnets, quantum
fluctuations about the magnetic ground states of CuFeO$_2$ can be
safely neglected due to the large $S=5/2$ spins.  Whereas geometric
frustration often leads to magnetic phases with non-collinear spins
and complex unit cells, magnetic anisotropy perpendicular to the
hexagonal planes in CuFeO$_2$ produces two different collinear
magnetic phases.  The $\uudd $ phase \cite{Mitsuda91, Mekata93}
sketched in Fig.1(a) is stable up to the field $B_{c1}\approx 7$ T .
Between $B_{c2}\approx 13.5$ T and $B_{c3}\approx 20$ T, another
collinear phase with a net moment of 1 $\mB$ per Fe$^{3+}$ ion
\cite{Mitsuda99, Terada06} has been assumed to resemble the $\uuudd$
phase shown in Fig.2 for type B stacking, with 5 Fe$^{3+}$ spins per
unit cell.  Incommensurate and non-collinear phases were identified
between $B_{c1}$ and $B_{c2}$ and above $B_{c3}$ \cite{Mitsuda99,
Terada06}.

Previous efforts to understand the collinear magnetic phases
\cite{Mekata93, Takagi95, Mitsuda99} and to estimate the exchange
and anisotropy parameters \cite{Ye07} of CuFeO$_2$ made the
simplifying assumption that the hexagonal layers were stacked
sequentially on top of each other.  We now demonstrate that an
accurate determination of the Heisenberg parameters must employ the
correct stacking of the hexagonal layers.  We also show that the
stability of a metamagnetic phase between $B_{c2}$ and $B_{c3}$
\cite{Mitsuda99, Terada06} strongly constrains those parameters.
Whereas earlier work \cite{Ye07} assuming a sequential stacking was
unable to explain the observed spin-wave (SW) frequencies of the
zero-field twins, realistic magnetic stackings are now used to
explain all features of the low-field collinear phase and to
identify the magnetic unit cell of the high-field collinear phase in
CuFeO$_2$.

The observation of collinear magnetic phases that are fully
polarized along the $\pm {\bf z}$ directions at low temperatures led
to the assumption \cite{Mekata93, Takagi95} that the Fe$^{3+}$ spins
were ``Ising-like."  However, measurements of the zero-field SW
frequencies \cite{Terada04, Ye07} plotted in Fig.1(b) reveal SW gaps
of only about 0.9 meV at wavevectors $(H,H,L=3/2)$ with $H=0.21$ and
0.29, on either side of the ordering wavevector $\vQ =
(1/4,1/4,3/2)$.  If the spins were truly ``Ising-like," then the SW
frequencies would be much higher and they would not exhibit a
significant dispersion along the $(0,0,L)$ direction \cite{Ye07,
Pet05} perpendicular to the hexagonal planes.  With little change in
wavevectors, the SW gaps are reduced either by an applied field
along the ${\bf z}$ axis or by the substitution of nonmagnetic
Al$^{3+}$ ions for Fe$^{3+}$.  Above the field $B_{c1}$
\cite{Mitsuda99, Terada06} or an Al concentration of about 1.6\%
\cite{Terada04}, the SW gaps vanish, the magnetic ground state
becomes non-collinear, and the crystals display multiferroic
behavior \cite{Kimura06, Kan07, Seki07}.

\begin{figure}
\includegraphics *[scale=0.42]{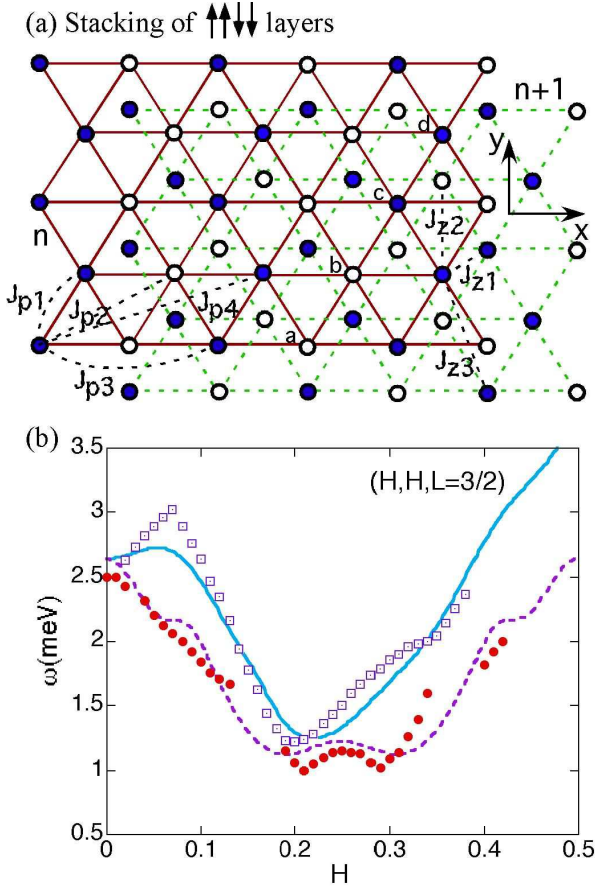}
\caption{
(a) The low-field spin configuration (up spins are empty and down
spins are filled circles) in each hexagonal plane, with the four
inequivalent spins $a $, $b $, $c $, and $d $.  Both the $n$ (solid)
and $n+1$ (dashed) layer are shown with the exchange parameters
indicated.  (b) The fit of the SW frequencies along the $(H,H,3/2)$
axis using the exchange and anisotropy parameters given in line
$iii$ of Table I.  Open squares give the frequencies of the twins
and solid circles the frequencies of the main SW branch with
ordering wavevector at $H=1/4$.
}
\end{figure}

Assuming that the hexagonal planes stack sequentially, we recently
fit \cite{Ye07} the SW frequencies of pure CuFeO$_2$ to the
predictions of the Heisenberg model 
\begin{equation}
\label{Ham}
H=-\frac{1}{2}\sum_{i\ne j}J_{ij}\vS_i \cdot \vS_j -D\sum_i
S_{iz}^2- 2\mu_B B\sum_i S_{iz},
\end{equation}
which includes single-ion anisotropy $D$ and a magnetic field $B$.
For ``Ising-like" spins, $D$ would be much greater than the exchange
parameters $J_{ij}$.  In a further simplification, we ignored the
very small ($< 0.4\%$) distortion of the hexagonal plane \cite{Ye06,
Terada06} below the N\'eel temperature.  While this distortion
breaks the symmetry between the $(H,H,0)$, $(H,0,0)$ and $(0,H,0)$
directions, thereby favoring the $\uudd $ phase with wavevector $\vQ
$ over its twins, it can produce only a very small change in the
exchange parameters and hence in the SW frequencies.  Despite these
simplifications, the SW dispersions evaluated along the $(H,H,3/2)$
and $(0,0,L)$ axis agree quite well with inelastic
neutron-scattering measurements \cite{Ye07}.  However, we were
unable to fit the frequency of the two twins with wavevectors
rotated $\pm \pi /3$ away from $\vQ $ in the $(H,K,3/2)$ plane.
Without attempting to fit the twins, we obtained the exchange and
anisotropy parameters given in line $i$ of Table I, where $J_{pm}$
or $J_{zm}$ are the $m$th nearest-neighbor exchange parameters
within each hexagonal plane or between adjacent planes.

\begin{table}
\begin{center}
\label{T1}
\caption{Heisenberg parameters of CuFeO$_2$ obtained from fits
of the zero-field SW frequencies.  Line $i$ assumes sequential stacking of the hexagonal layers \cite{Ye07}, 
$ii$ and $iii$ use the realistic stacking in Fig.1(a) while $iii$ also constrains the parameters to stabilize the collinear phase
between $B_{c2}$ and $B_{c3}$. Exchange and anisotropy parameters are in meV;  $T_N^{MF}$ is in K.}
\begin{tabular}{p{0.2in} p{0.3in} p{0.3in} p{0.3in} p{0.3in} p{0.3in} p{0.3in} p{0.3in} p{0.3in} p{0.3in}} \hline
fit &  $\, \, J_{p1}$ & $\, \, J_{p2}$  & $\, \, J_{p3}$ & $\, \, J_{p4}$ & $\, \, J_{z1}$ & $\, \, J_{z2}$ & $\, \, J_{z3}$ & $\, \, D $ & $T_N^{MF}$ \\ \hline
$i$ & -0.46 & -0.20 & -0.26  &   & -0.13 & $\,\, $0.00 &  & 0.07   &46\\ 
$ii$ & -0.75 & -0.17 & -0.10 & $\,\,$0.01& -0.51 & -0.19 & -0.06 & 0.14 &65\\
$iii$ & -0.23 & -0.12 & -0.16 & $\,\, $0.00 & -0.06 & $\,\, $0.07 & -0.05 & 0.22 &25 \\
\hline
\end{tabular}
\end{center}
\end{table}

To better understand the metamagnetic phase between $B_{c2}$ and
$B_{c3}$, we have recalculated the SW frequencies of the $\uudd$
phase below $B_{c1}$ using the realistic magnetic stacking of the
hexagonal layers shown in Fig.1(a).  All other stackings of the
$\uudd $ layers have higher coupling energies.  Because spins $a$,
$b$, $c$, or $d$ experience the same local environment on every
layer, the magnetic unit cell still contains only 4 sublattices
(SLs).  The first few exchange pathways $J_{pm}$ and $J_{zm}$ are
indicated in Fig.1(a).   

The SW frequencies are evaluated using a Holstein-Primakoff (HP)
$1/S$ expansion about the classical limit.  On the spin-up $a$ and
$b$ sites, we replace $S_{iz} =S-\alpha_i^{\dagger }\alpha_i$,
$S_i^+ = S_{ix}+iS_{iy}=\sqrt{2S}\alpha_i$, and $S_i^-=
S_{ix}-iS_{iy} =\sqrt{2S}\alpha_i^{\dagger }$ ($\alpha_i = a_i$ or
$b_i$).  On the the spin-down $c$ and $d$ sites, we replace $S_{iz}
=-S+\gamma_i^{\dagger }\gamma_i$, $S_i^+=\sqrt{2S}\gamma_i^{\dagger
}$, and $S_i^- =\sqrt{2S}\gamma_i$ ($\gamma_i = c_i$ or $d_i$).
The SW frequencies $\omega_{\vk }$ at wavevector $\vk $ are then
obtained by solving the equations-of-motion for the vectors
$\vv_{\vk }= ( a_{\vk } , b_{\vk }, c_{\vk }^{\dagger }, d_{\vk
}^{\dagger })$ and $\vv_{\vk }^{\dagger }$.  The equation-of-motion
for $\vv_{\vk }$ may be written in terms of the 4 x 4 matrix $\uM
(\vk )$ as $id\vv_{\vk }/dt =-[H,\vv_{\vk }]=\uM (\vk )\vv_{\vk }$
with SW frequencies given by the condition $\Det (\uM (\vk
)-\omega_{\vk }\underline{I})=0$.  Only positive frequencies
$\omega_{\vk } \ge 0$ are retained.

As expected for a collinear antiferromagnet and shown schematically
for any wavevector in Fig.3(a), each of the SW branches is linearly
split by a magnetic field.  The lowest SW frequency with wavevector
$(0.21,0.21,1.5)$ or $(0.29,0.29,1.5)$ will vanish at the field 0.9
meV$/ 2\mu_B \approx 7.7$ T, which is slightly larger than $B_{c1}$.
With the correct stacking of the hexagonal layers, the parameters in
line $ii$ of Table I are obtained by fitting the SW frequencies of
the main branches along the $(H,H,3/2)$ and $(0,0,L)$ axis as well
as the SW frequencies of the twins evaluated along $(H,H,3/2)$. 

\begin{figure}
\includegraphics *[scale=0.40]{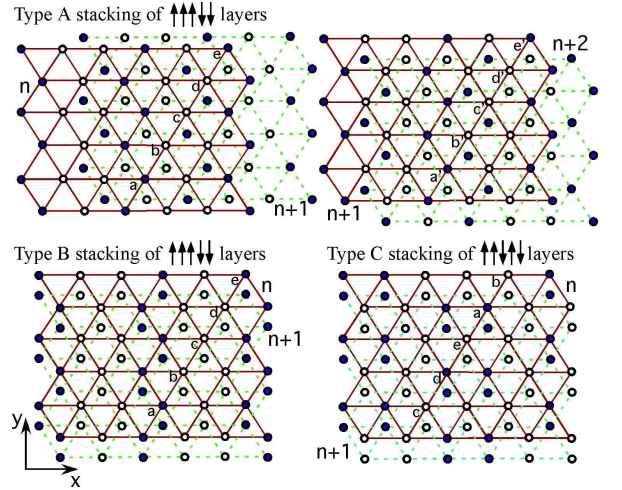}
\caption{
Three types of magnetic stacking that satisfy the conditions for
local stability of the metamagnetic phase.  In type A stacking, the
stacking patterns on the left and right alternate.
}
\end{figure}

There are several possible collinear metamagnetic phases with a net
moment of 1 $\mB $ per Fe$^{3+}$ ion and with elastic peaks at
wavevectors $(m/5,m/5,0)$ in the $L=0$ basal plane \cite{Mitsuda99}.
Two configurations are possible in each hexagonal plane:  the
$\uuudd $ pattern sketched in the lower left of Fig.2 and the
$\uudud$ pattern sketched in the lower right.   Depending on the
stacking, the magnetic unit cell of the metamagnetic phase may
contain either 5 or 10 magnetic ions.   For example, type A stacking
of $\uuudd $ layers in Fig.2 contains 10 SLs, while type B stacking
of $\uuudd $ layers and type C stacking of $\uudud$ layers contain 5
SLs.  In type A stacking, the local environments of spin $a$ on
layer $n$ and spin $a'$ on layer $n+1$ are different:  $a$ is
coupled by $J_{z1}$ to three up spins on layer $n+1$ while $a'$ is
coupled by $J_{z1}$ to two up spins and one down spin on layer
$n+2$.  In types B and C stacking, the spins in layer $n+1$ are
obtained from those in layer $n$ by the displacement $-\sqrt{3}{\bf
y}/3$.  For a 5 or 10 SL stacking, the matrix $\uM (\vk )$ that
enters the equations-of-motion for the SW frequencies is  5 or 10
dimensional and the 5 or 10 SW branches $s$ must be solved
numerically for every $\vk $. 

Two conditions must be satisfied for the local stability of a
metamagnetic phase.  First, the SW frequencies $\omega_{\vk }^{(s)}$
must all be real.  This condition is independent of the magnetic
field, which only shifts the frequencies by $\pm 2\mB B$, and is not
always satisfied because $\uM (\vk )$ is not Hermitian.  Second, the
SW weights $W^{(s)}_{\vk }$ that appear as coefficients of the delta
functions in the spin-spin correlation function 
\begin{eqnarray}
S(\vk ,\omega )&=&\ds\frac{1}{N} \int dt \,e^{-i\omega t} \sum_{i,j} e^{i\vk \cdot (\vR_j-\vR_i)}\Bigl\{ 
\langle S_i^+ S_j^- (t)\rangle 
\nonumber \\&
+&\langle S_i^- S_j^+ (t)\rangle \Bigr\}  = \sum_s W^{(s)}_{\vk} \delta (\omega -\omega_{\vk }^{(s)})
\end{eqnarray}
must all be positive.  Those weights are most easily evaluated by
expanding $S(\vk ,\omega )$ within the HP formalism and then solving
the equations-of-motion for the spin Green's functions.  An
equivalent but much easier way to guarantee that the weights
$W^{(s)}_{\vk }$ are positive is to examine the field-dependence of
the SW frequencies.  For a stable 5 SL collinear phase, 3 of the 5
SW modes must linearly increase with field while 2 must linearly
decrease, as shown in Fig.3(a).  For a stable 10 SL collinear phase,
6 of the 10 SW modes must linearly increase and 4 must linearly
decrease with field.  If this condition is violated for any $\vk $,
then some of the weights $W^{(s)}_{\vk }$ will be negative and the
phase will be unstable.

Unfortunately, the exchange and anisotropy parameters given by lines
$i$ and $ii$ of Table I do not satisfy both conditions for the local
stability of any possible stacking of $\uuudd $ or $\uudud $ layers
between the fields $B_{c2}$ and $B_{c3}$.   In other words, fits to
the SW frequencies of the zero-field $\uudd $ phase are inconsistent
with the existence of a collinear metamagnetic phase.

\begin{figure}
\includegraphics *[scale=0.43]{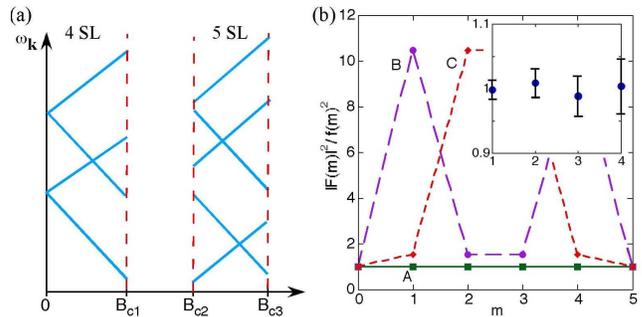}
\caption{
(a) Schematic field dependence of the SW frequencies
$\omega^{(s)}_{\vk }$ in 4 or 5 SL phases.  (b) The predicted
elastic intensities $\vert F(m)\vert^2$ normalized by the Fe$^{3+}$
form factor $f(m)^2$ versus $m$ ($H=m/5$ along the $(H,H,0)$ axis)
for stackings A, B, and C of the metamagnetic phases.  Inset are the
experimental, normalized intensities versus $m$.
}
\end{figure}

This inconsistency may be eliminated by fitting the zero-field SW
frequencies of the $\uudd $ phase while simultaneously constraining
the exchange and anisotropy parameters to stabilize a metamagnetic
phase between $B_{c2}$ and $B_{c3}$.   We emphasize that this
constraint utilizes only the observed stability of the metamagnetic
phase over a range of magnetic fields and not the measured SW
frequencies of that phase.  The three phases shown in Fig.2 are the
only ones that satisfy both conditions for local stability when the
exchange and anisotropy parameters are obtained from constrained
zero-field fits of the SW frequencies.  

To determine which of these three phases is observed, we evaluate
the magnetic structure factor $F(m)$ for the elastic peaks in the
$L=0$ basal plane at wavevectors $(H,H,0)$ with $H=m/5$:  
\begin{equation}
F^A(m)=f(m)e^{2\pi iH},
\end{equation}
\begin{equation}
F^B(m)=f(m)e^{2\pi iH}\Bigl\{ 2i \sin 2\pi H +2i \sin 4\pi H +1\Bigr\},
\end{equation}
\begin{equation}
F^C(m)=f(m)e^{2\pi iH}\Bigl\{ 2\cos 2\pi H -2i \sin 4\pi H -1\Bigr\},
\end{equation}
where $f(m)$ is the magnetic form factor of each Fe$^{3+}$ ion.
Notice that $F(0)=f(0)$ for all three possible phases.  The
normalized intensities $\vert F(m)\vert^2/f(m)^2$ are plotted versus
$m$ in Fig.3(b).  When the magnetic moments of 6 adjacent layers are
summed, stacking A produces the pattern 0$\uparrow$000 along ${\bf
x}$ so that $\vert F^A(m)\vert^2/f(m)^2=1$ is constant.  If the
stacking pattern on the left or right top panel of Fig.2 were
continued indefinitely rather than alternating, then the resulting
phase would have no elastic peaks in the $L=0$ basal plane.  The
layer sums of stackings B or C produce $\uuudd $ or $\uudud$
patterns along ${\bf x}$, causing $\vert F(m)\vert^2 /f(m)^2$ to
change by a factor of 10.5 as $m$ increases from 1 to 2.

For comparison, the experimental results \cite{Kan07} for the
elastic intensities are plotted in the inset to Fig.3(b).  The
normalized intensity $\vert F(m)\vert^2/f(m)^2$ for $m=1$ through 4
is constant to within about 1\%.   Therefore, only type A stacking
of $\uuudd$ layers with a 10 SL unit cell is possible.

The exchange and anisotropy parameters associated with stacking A
are given on line $iii$ of Table I.  As in our original fits
\cite{Ye07}, $\vert J_{p3}\vert  > \vert J_{p2}\vert $ but $J_{p4}$
is negligible.  Since $J_{z3}$ is comparable to $J_{z1}$, even
longer-ranged interactions between neighboring planes might exist.
All of the interactions $J_{zm}$ between adjacent planes are much
smaller in magnitude than the interactions $J_{pm}$ ($m < 4)$ within
a plane.  Using these parameters, the fits of the main and twin SW
branches are plotted along the $(H,H,3/2)$ axis in Fig.1(b).

Constraining the fits of the zero-field SW frequencies to produce a
stable metamagnetic phase has a substantial effect on the exchange
and anisotropy parameters.  For example, $J_{p1}$ is reduced by
about 70\% from line $ii$ to line $iii$ of Table I.  While a wide
range of parameters can provide reasonable fits to the zero-field SW
data,  demanding that a metamagnetic phase is stabilized between
$B_{c2}$ and $B_{c3}$ considerably narrows the possible range of
those parameters.  Also notice that the mean-field transition
temperature $T_N^{MF}$ listed in line $iii$ of Table I is much
closer to the measured transition temperature of 14 K
\cite{Mitsuda99} between partially-disordered and paramagnetic
phases than the transition temperatures of the unconstrained fits in
lines $i$ and $ii$.  Of all three fits, line $iii$ produces a
crystal-field environment that is most ``Ising-like," with the
anisotropy $D$ about the same size as the nearest-neighbor exchange
$J_{p1}$.  The difference between the parameters in lines $i$, $ii$,
and $iii$ of Table I underscores the danger of using even an
extensive set of SW measurements for a single magnetic phase to fix
the parameters of a Heisenberg model.

Surprisingly, the high-field collinear phase is the 10 SL phase
sketched in Fig.2 rather than the 5 SL phase that had been
previously assumed \cite{Mitsuda99, Terada06}.  Because it remains
locally stable up to about 34.5 T (very close to the critical field
$B_{c4}$ measured by Terada {\em et al.} \cite{Terada06}), the
disappearance of the 10 SL $\uuudd$ phase at $B_{c3}$ probably
occurs at a first-order transition between collinear and
non-collinear phases.  That appears to be the case for the $\uudd $
phase, since $B_{c1}$ is lower than the 7.7 T field where the SW gap
would vanish and the $\uudd $ phase would become locally unstable.
The 10 SL $\uuudd $ phase remains locally stable only down to
$B_{c2}$, where the frequency of a SW mode vanishes.  

Our work demonstrates that the stacking of the hexagonal planes and
the stability of a metamagnetic phase play crucial roles in
determining the exchange and anisotropy parameters of a frustrated
antiferromagnet.  By constraining the fitting parameters at zero
field, we have been able to identify the magnetic unit cell of the
collinear metamagnetic phase in CuFeO$_2$.  Constrained zero-field
fits may prove to be a powerful technique for other systems as well.

We would like to acknowledge helpful conversations with Satoshi
Okamoto.  This research was sponsored by the Laboratory Directed
Research and Development Program of Oak Ridge National Laboratory,
managed by UT-Battelle, LLC for the U. S. Department of Energy under
Contract No. DE-AC05-00OR22725 and by the Division of Materials
Science and Engineering and the Division of Scientific User
Facilities of the U.S. DOE.


\begin{references}

\suppressfloats

\bibitem{Diep04} See, for example, {\it Frustrated Spin Systems}
(World Scientific, New Jersey, 2004), edited by H.T. Diep.

\bibitem{Mitsuda91} S. Mitsuda, H. Yoshizawa, N. Yaguchi, and M.
Mekata, {\it J. Phys. Soc. Jpn.} {\bf 60}, 1885 (1991).

\bibitem{Mekata93} M. Mekata, N. Yaguchi, T. Takagi, T. Sugino, S.
Mitsuda, H. Yoshizawa, N. Hosoito, and T. Shinjo, {\it J. Phys. Soc.
Jpn.} {\bf 12}, 4474 (1993).

\bibitem{Mitsuda99} S. Mitsuda, M.Mase, T. Uno, H. Kitazawa, and
H.A. Katori, {\it J. Phys.  Chem. Sol.} {\bf 60}, 1249 (1999);  S.
Mitsuda, M. Mase, K. Prokes, H. Kitazawa, and H.A. Katori, {\it J.
Phys. Soc. Jpn.} {\bf 69}, 3513 (2000).

\bibitem{Terada06} N. Terada, Y. Narumi, K. Katsumata, T. Yamamoto,
U. Staub, K. Kindo, M. Hagiwara, Y. Tanaka, A. Kikkawa, H. Toyokawa,
T. Fukui, R. Kanmuri, T. Ishikawa, and H. Kitamura, {\it Phys. Rev.
B} {\bf 74}, 180404(R) (2006);  N. Terada, Y. Narumi, Y. Sawai, K.
Katsumata, U. Staub, Y. Tanaka, A. Kikkawa, T. Fukui, K. Kindo, T.
Yamamoto, R. Kanmuri, M. Hagiwara, H. Toyokawa, T.  Ishikawa, and H.
Kitamura, {\it Phys. Rev. B} {\bf 75}, 224411 (2007).

\bibitem{Takagi95} T. Takagi and M. Mekata, {\it J. Phys. Soc. Jpn.}
{\bf 64}, 4609 (1995).

\bibitem{Ye07} F. Ye, J.A. Fernandez-Baca, R.S. Fishman, Y. Ren,
H.J. Kang, Y. Qiu, and T. Kimura, {\it Phys. Rev. Lett.} {\bf 99},
157201 (2007);  R.S. Fishman, {\it J. Appl. Phys.} {\bf 103}, 07B109
(2008).

\bibitem{Terada04} N. Terada, S. Mitsuda, Y. Oohara, H. Yoshizawa,
and H. Takei, {\it J. Magn. Magn. Mat.} {\bf 272-276}, e997 (2004);
N. Terada, S. Mitsuda, K. Prokes, O. Suzuki, H. Kitazawa, and H.A.
Katori, {\it Phys. Rev. B} {\bf 70}, 174412 (2004); N. Terada, S.
Mitsuda, T. Fujii, and D. Petitgrand, {\it J. Phys.: Cond. Mat.}
{\bf 19}, 145241 (2007).

\bibitem{Pet05} O.A. Petrenko, M.R. Lees, G. Balakrishnan, S. de
Brion, and G. Chouteau, {\it J. Phys.: Cond. Mat.} {\bf 17}, 2741
(2005).

\bibitem{Kan07} S. Kanetsuki, S. Mitsuda, T. Nakajima, D. Anazawa,
H.A. Katori, and K. Prokes, {\it J. Phys.: Cond. Mat.} {\bf 19},
145244 (2007).

\bibitem{Kimura06} T. Kimura, J.C. Lashley, and A.P. Ramirez, {\it
Phys. Rev. B} {\bf 73}, 220401(R) (2006).

\bibitem{Seki07} S. Seki, Y. Yamasaki, Y. Shiomi, S. Iguchi, Y.
Onose, and Y. Tokura, {\it Phys. Rev. B} {\bf 75}, 100403(R) (2007).

\bibitem{Ye06} F. Ye, Y. Ren, Q. Huang, J.A. Fernandez-Baca, P. Dai,
J.W. Lynn, and T. Kimura, {\it Phys. Rev. B} {\bf 73}, 220404(R)
(2006).

\end{references}
\end{document}